\documentclass{article}
\usepackage{epsf,amssymb}

\begin{document}
\title{Gravitational Solitons and Monodromy Transform
Approach to Solution of Integrable Reductions of Einstein Equations}
\author{G.A.~Alekseev\\
{\it Steklov Mathematical Institute, Russian Ac. Sci.,}\\
{\it Gubkina 8, 117966, GSP - 1, Moscow, Russia}
\\{\it E-mail: G.A.Alekseev@mi.ras.ru}}
\date{November 15, 1999}
\maketitle

\begin{abstract}
In this paper the well known Belinskii and Zakharov soliton
generating transformations of the solution space of vacuum Einstein
equations with two-dimensional Abelian groups of isometries are
considered in the context of the so called "monodromy transform
approach", which provides some general base for the study of various
integrable space - time symmetry reductions of Einstein equations.
Similarly to the scattering data used in the known spectral
transform, in this approach the monodromy data for solution of
associated linear system characterize completely any solution of the
reduced Einstein equations, and many physical and geometrical
properties of the solutions can be expressed directly in terms of the
analytical structure on the spectral plane of the corresponding
monodromy data functions. The Belinskii and Zakharov vacuum soliton
generating transformations can be expressed in explicit form (without
specification of the background solution) as simple
(linear-fractional) transformations of the corresponding monodromy
data functions with coefficients, polynomial in spectral parameter.
This allows to determine many physical parameters of the generating
soliton solutions without (or before) calculation of all components of
the solutions.  The similar characterization for electrovacuum
soliton generating transformations is also presented.  \end{abstract}

\section{Introduction}

The existence of very rich, integrable structure of Einstein
equations, at least for space-times with two-dimensional Abelian
isometry groups, have been conjectured by different authors many
years ago, but the real discoveries of beautiful integrability
properties of these equations and effective procedures for
construction of their solutions have been started more than twenty
years ago in the papers of Belinskii and Zakharov \cite{BZ:1978,
BZ:1979}. In these papers the inverse scattering methods have been
developed for Einstein equations for vacuum gravitational fields. In
particular, the solution of the entire problem had been reduced to
some matrix Riemann - Hilbert problem and soliton generating
(dressing) technique had been suggested for calculation of vacuum
gravitational solitons on an arbitrary chosen (vacuum) background.
Later numerous investigations of integrable reductions of Einstein
equations (for vacuum or in the presence of electromagnetic and some
other matter fields) have been made by many authors, using different
powerful ideas of the modern theory of completely integrable systems.
A number of different more or less general approaches
were developed and many other interesting results were
obtained.\footnote{Avoiding a detail citation, we refer the readers
to the references in a few papers cited here, but mainly -- to a
large and very useful F.J.Ernst's collection of related references
and abstracts, accessible throw {\sf http://pages.slic.com/gravity}.}

Here we consider Belinskii and Zakharov vacuum
soliton generating transformations in the context of so called
"monodromy transform" approach. This approach, developed by the
author in \cite{GA:1985} - \cite{GA:1999}, provides some general
base for the analysis of all known integrable reductions of Einstein
equations. In this approach any local solution of
reduced Einstein equations is characterized by a set of monodromy
data of the fundamental solution of some associated spectral
problem.\footnote{Though the similar constructions surely can be
realized for any of the gauge equivalent null curvature
representations of the field equations, there were used
in this approach a spectral plane $w$, which is covered twice by the
Belinskii and Zakharov spectral plane $\lambda$, and different
spectral problem (first constructed by Kinnersley and
Chitre \cite{KCIII:1978} in the context of their group - theoretic
approach), which is gauge equivalent to the Belinskii and Zakharov
one. The reason for the first difference is the absence of
differentiation in the transformed linear system with respect to a
spectral parameter $w$, and for the second one - the simplest
monodromy properties of solutions on this spectral plane.} The direct
and inverse problems of such monodromy transform possess unambiguous
solutions.\footnote{The linear singular integral equation with a
scalar kernel, which solves the inverse problem of this monodromy
transform, as well as its equivalent regularization -- a (quasi-)
Fredholm equation of the second kind have been presented in
\cite{GA:1985, GA:1988} and \cite{GA:1999} respectively.} It is
remarkable, that many physical and geometrical properties of the
solutions can be expressed directly in terms of the analytical
structure of these monodromy data on the spectral plane. A lot of the
known physically interesting solutions possess very simple,
rational structures of these monodromy data functions. As it will be
shown below, it is also convenient to characterize the Belinskii and
Zakharov vacuum soliton generating procedure in terms of the
corresponding transformation of the monodromy data of the background
solution into the monodromy data of the (multi-) soliton solution.
These transformations possess an explicit and
remarkably simple, linear-fractional form with coefficients,
polynomial in the spectral parameter of the orders less or equal
to a number of solitons. This allows to calculate various physical
parameters of the generating soliton solutions without (or
before) their complete calculation. A generalization
of these transformations for the case of electrovacuum solitons,
found in \cite{GA:1980}, is also presented.

\section{Belinskii and Zakharov gravitational solitons}
The dynamical part of vacuum Einstein equations for the space-time
metrics
\begin{equation}\label{metrics}
ds^2=f(x^1,x^2)\eta_{\mu\nu} dx^\mu
dx^\nu+g_{ab}(x^1, x^2) dx^a dx^b\end{equation}
where $\mu,\nu=1,2$; $a,b=3,4$; $f>0$ and
$\eta_{\mu\nu}=\mbox{diag\,}\{\epsilon_1,\epsilon_2\}$ with
$\epsilon_1=\pm 1$, $\epsilon_2=\pm 1$, can be reduced to the
form (used in \cite{BZ:1978, BZ:1979} in a bit different
notation):
\begin{equation}\label{BZequations}
\left\{
\begin{array}{l}
\eta^{\mu\nu}\partial_\mu (\alpha\,\partial_\nu{\bf g}\cdot
{\bf g}^{-1})=0\\[1ex]
{\bf g}^T={\bf g},\quad \det {\bf g}=
\epsilon\alpha^2\end{array}\hskip1ex
\right\Vert\hskip1ex
\left.
\begin{array}{l}
\epsilon\equiv-\epsilon_1\epsilon_2\\[1ex]
\partial_\mu\partial_\nu\alpha=0
\end{array} \hskip1ex\right\Vert\hskip1ex
\begin{array}{l}
\beta: \\[1ex]\phantom{\beta:}
\end{array}
\begin{array}{l}
\partial_1\beta=\epsilon_1\partial_2\alpha,\\[1ex]
\partial_2\beta=-\epsilon_2\partial_1\alpha
\end{array}
\end{equation}
where ${\bf g}=\Vert g_{ab}\Vert$ is a real symmetric
 $2\times 2$ - matrix, and the given above definition of
$\epsilon$ is implied by the Lorentz signature of the metric
(\ref{metrics}). Therefore, $\epsilon=1$ corresponds to a hyperbolic
case and $\epsilon=-1$ is the elliptic case. Note, that the linear
"harmonic" equation for $\alpha$, which follows immediately from the
trace of the equation for ${\bf g}$ in (\ref{BZequations}),
provides the existence of the defined above function $\beta$,
"harmonically" conjugated to $\alpha$. It is convenient to use farther
these geometrically defined functions as new coordinates in the
linear combinations $\xi=\beta+j\alpha$, $\eta=\beta-j\alpha$, where
$j=1$ for $\epsilon=1$ and $j=i$ for $\epsilon=-1$.

In these notation the Belinskii and Zakharov spectral problem reads
as
\begin{equation}\label{BZlinsys}
\left\{\begin{array}{l}
D_\xi{\bf \Psi}_{\scriptscriptstyle{BZ}}=\displaystyle{{\mathbb
V}_\xi\over\lambda-j\alpha}{\bf\Psi}_{\scriptscriptstyle{BZ}}\\[2ex]
D_\eta{\bf \Psi}_{\scriptscriptstyle{BZ}}=\displaystyle{{\mathbb V}_\eta
\over\lambda+j\alpha}{\bf\Psi}_{\scriptscriptstyle{BZ}}\end{array} \hskip1ex
\right\vert \hskip0.5ex \left.\begin{array}{l}
D_\xi=\partial_\xi-\displaystyle{\lambda\over\lambda-j\alpha}
\displaystyle{\partial\over\partial\lambda} \\[2ex]
D_\eta=\partial_\eta-\displaystyle{\lambda\over\lambda+j\alpha}
\displaystyle{\partial\over\partial\lambda}
\end{array}
\hskip1ex \right\vert\hskip0.5ex
\begin{array}{l}
{\mathbb V}_\xi=-j\alpha\,\partial_\xi{\bf g}\cdot
{\bf g}^{-1}\\[3ex]
{\mathbb V}_\eta=j\alpha\,\partial_\eta {\bf g}\cdot {\bf g}^{-1}
\end{array}
\end{equation}
with additional "reduction" conditions imposed on the
solutions of (\ref{BZlinsys}):
\begin{equation}\label{BZint}
{\bf \Psi}^T_{\scriptscriptstyle{BZ}}(\lambda\to
\displaystyle{\epsilon\alpha^2\over\lambda})\cdot{\bf
g}^{-1}\cdot{\bf\Psi}_{\scriptscriptstyle{BZ}}=
{\bf K}_{\scriptscriptstyle{BZ}}(w),
\quad{\bf K}^T_{\scriptscriptstyle{BZ}}= {\bf
K}_{\scriptscriptstyle{BZ}}, \quad D_\xi w=D_\eta w=0
\end{equation}

For the construction of vacuum solitons, based on
the form (\ref{BZlinsys}), (\ref{BZint}) of the spectral problem, the
dressing transformation  ${\bf
\Psi}_{\scriptscriptstyle{BZ}}(\xi,\eta,\lambda)={\bf
\chi}_{\scriptscriptstyle{BZ}}(\xi,\eta,\lambda)\cdot {\stackrel o {\bf
 \Psi}_{\scriptscriptstyle{BZ}}}(\xi,\eta,\lambda) $ was used in
\cite{BZ:1978, BZ:1979} with the soliton ansatz\footnote{In the case
$\epsilon=1$ the number of solitons (poles) can be odd as well, but
we shall not consider this case here.}
\begin{equation}\label{lamsol}
{\bf \chi}_{\scriptscriptstyle{BZ}}={\mathbb
I}+\sum\limits_{k=1}^{2 N}\displaystyle{{\mathbb
R}_k(\xi,\eta)\over\lambda-\mu_k(\xi,\eta)},\qquad
{\bf \chi}_{\scriptscriptstyle{BZ}}^{-1}={\mathbb
I}+\sum\limits_{k=1}^{2 N}\displaystyle{{\mathbb
S}_k(\xi,\eta)\over\lambda-\nu_k(\xi,\eta)}\end{equation}
where ${\mathbb I}$ is the identity matrix, the $2\times 2$-matrix
residues at the poles as well as the pole trajectories $\mu_k$ and
$\nu_k$ are real or constitute complex conjugated pairs. Then
in \cite{BZ:1978, BZ:1979} all
constraints on these functions, which follow from (\ref{BZlinsys}),
had been successfully solved and all metric components of the
generating soliton solutions have been explicitly expressed in
terms of a number of integration constants and the matrix ${\stackrel
o {\bf \Psi}_{\scriptscriptstyle{BZ}}}(\xi,\eta,\lambda)$,
characterizing arbitrarily chosen
"background" solution. (See also \cite{GA:1981} for more compact,
determinant form of these soliton solutions).

\section{Vacuum $w$ - solitons.}

The function $w$, which was used as the new spectral parameter in
the mentioned above monodromy transform approach, have been
introduced also in \cite{BZ:1978, BZ:1979}, as a solution of the
equations $D_\xi w=D_\eta w=0$, such that
$2 w=\lambda+2\beta +\displaystyle{\epsilon\alpha^2 /
\lambda}.$ Thus, the spectral plane $\lambda$ covers twice the
spectral plane $w$. The  Kinnersley - Chitre-like linear system
\cite{KCIII:1978} which substitute  in our approach the linear
system (\ref{BZlinsys}), and the gauge transformation between these
systems in our notation are
\begin{equation}\label{UVvlinsys}
\left\{\begin{array}{l}
\partial_\xi{\bf \Psi}=\displaystyle{{\bf U}(\xi,\eta)\over 2 i
(w-\xi)}{\bf\Psi}\\[2ex] \partial_\eta{\bf \Psi}=\displaystyle{{\bf
V}(\xi,\eta)\over 2 i (w-\eta)}{\bf\Psi}\end{array}
\quad\right\Vert\quad
\begin{array}{l}
{\bf \Psi}_{\scriptscriptstyle{BZ}}(\xi,\eta,\lambda)={\bf A}_\otimes \cdot{\bf
\Psi}(\xi,\eta,w)\\[3ex]
{\bf A}_\otimes={\bf g}\cdot {\mathbb \varepsilon}+
i\lambda\,{\mathbb I},\quad
{\mathbb \varepsilon}=\left(\hskip-1ex\begin{array}{rr}
0&1\\-1&0\end{array}\right) \end{array} \end{equation}
The following two groups of conditions constitute a spectral problem,
based on the linear system (\ref{UVvlinsys}) and equivalent to the
reduced vacuum equations (\ref{BZequations})
\cite{GA:1980, GA:1988}:
\begin{eqnarray}
&\left\{\begin{array}{l}
2 i (w-\xi)\partial_\xi{\bf \Psi}={\bf
U}(\xi,\eta)\cdot{\bf\Psi}\\[1ex] 2 i (w-\eta)\partial_\eta{\bf
\Psi}={\bf V}(\xi,\eta)\cdot{\bf\Psi} \end{array} \hskip2ex
\right\Vert\hskip2ex \begin{array}{l} \mbox{rank\,}{\bf U}=1,\quad
\mbox{tr\,}{\bf U}= i, \\[1ex] \mbox{rank\,}{\bf V}=1,\quad
\mbox{tr\,}{\bf V}= i, \end{array}\label{UVvaclinsys}
\\[1ex]
&\left\{\begin{array}{l}
{\bf \Psi}^\dagger\cdot{\bf W}\cdot{\bf \Psi}={\bf K}(w)\\[1ex]
{\bf K}^\dagger(w)={\bf K}(w)
\end{array}\hskip6ex\right\Vert\hskip2ex  \displaystyle{\partial
{\bf W} \over \partial w} = 4 i {\mathbb\varepsilon},\quad
{\mathbb \varepsilon}
= \left(\hskip-1ex\begin{array}{rr} 0&1\\-1&0\end{array}\right)
\label{Wvacint} \end{eqnarray}
For the construction of the Belinskii and Zakharov vacuum solitons,
based on the spectral problem (\ref{UVvaclinsys}), (\ref{Wvacint}), we use
similar dressing transformation ${\bf \Psi}(\xi,\eta,w)={\bf
\chi}(\xi,\eta,w)\cdot {\stackrel o {\bf \Psi}}(\xi,\eta,w)$ with
a bit different, than in (\ref{lamsol}), soliton ansatz
\footnote{For the case of real poles an alternative
construction of the Belinskii and Zakharov solitons
in the context of the spectral problem (\ref{UVvaclinsys}),
(\ref{Wvacint}), which corresponds to the choice $X(w)=1$ in
(\ref{wsolvac}) and ${\bf K}(w)=\prod_{k=1}^N
\left({w-\widetilde{w}_k\over w-w_k}\right){\stackrel o {\bf K}}(w)$,
had been presented in \cite{SM_JBG:1999}.}
\begin{equation}\label{wsolvac}
{\bf \chi}=X(w)\left({\bf
I}+\sum\limits_{k=1}^{N}\displaystyle{{\bf R}_k(\xi,\eta)\over
w-w_k}\right),\qquad
{\bf \chi}^{-1}=\displaystyle{1\over X(w)}\left({\bf
I}+\sum\limits_{k=1}^{N}\displaystyle{{\bf
S}_k(\xi,\eta)\over w-\widetilde{w}_k}\right),
\end{equation}
where $X(w)=\prod_{k=1}^N \left({w-w_k\over
w-\widetilde{w}_k}\right)^{1\over 2}$, the gauge condition ${\bf
K}(w)={\stackrel o {\bf K}}(w)$ is used,
the constant pole locations $w_k$ and $\widetilde{w}_k$ for each
$k=1,2,\ldots, N$ are the pairs of different real or complex
conjugated constants,
\footnote{It is useful to note, that the number
$N$ of $w$-poles (solitons) in (\ref{wsolvac}) corresponds to $2 N$
solitons ($\lambda$-poles) in (\ref{lamsol}); each
real pole with $w_k\ne \widetilde{w}_k$ is equivalent to a pair of
real $\lambda$ - poles in (\ref{lamsol}), while each of the complex
$w$-poles with $\widetilde{w}_k=\overline{w_k}$ correspond to a pair
of complex conjugated to each other $\lambda$-poles in
(\ref{lamsol}).} and the values, having "o" overhead, correspond to
the background solution. Again, similarly to the original Belinskii
and Zakharov construction of $\lambda$-solitons, all constraints,
which follow from (\ref{UVvaclinsys}), (\ref{Wvacint}) for the matrix
residues in (\ref{wsolvac}) can be solved \cite{GA:1980, GA:1988}, and
for any choice of the background solution the generating $N$-soliton
solution can be expressed in terms of the background ${\stackrel o
{\bf \Psi}}(\xi,\eta,w)$ and a set of integration constants (four
real constants per each of the poles $w_k$).

\section{Electrovacuum $w$ - solitons}
One of the important features of the  spectral problem
(\ref{UVvaclinsys}), (\ref{Wvacint}) is that very small changes of
its structure lead to a matrix problem, equivalent to the
space-time symmetry reduced electrovacuum Einstein - Maxwell
equations \cite{GA:1980, GA:1988, GA:1999}:  \begin{eqnarray}
&\left\{\begin{array}{l}
2 i (w-\xi)\partial_\xi{\bf \Psi}={\bf
U}(\xi,\eta)\cdot{\bf\Psi}\\[1ex] 2 i (w-\eta)\partial_\eta{\bf
\Psi}={\bf V}(\xi,\eta)\cdot{\bf\Psi} \end{array} \hskip2ex
\right\Vert\hskip2ex \begin{array}{l} \mbox{rank\,}{\bf U}=1,\quad
\mbox{tr\,}{\bf U}= i, \\[1ex] \mbox{rank\,}{\bf V}=1,\quad
\mbox{tr\,}{\bf V}= i, \end{array}\quad\label{UVlinsys}
\\[1ex]
&\left\{\begin{array}{l}
{\bf \Psi}^\dagger\cdot{\bf W}\cdot{\bf \Psi}={\bf K}(w)\\[1ex]
{\bf K}^\dagger(w)={\bf K}(w)
\end{array}\hskip1ex\right\Vert\hskip2ex
\displaystyle{\partial
{\bf W} \over \partial w} = 4 i \left(\hskip-1ex\begin{array}{rrr}
0&1&0\\-1&0&0\\0&0&0\end{array}\right),\hskip2ex
{\bf W}^{55}=1
 \label{Wint} \end{eqnarray}
where the unknown matrix variables ${\bf \Psi}(\xi,\eta,w)$, ${\bf
U}(\xi,\eta)$, ${\bf V}(\xi,\eta)$ and ${\bf W}(\xi,\eta,w)$ are now
$3\times 3$ - matrices and ${\bf W}^{55}$ denotes the lower
right component of $3\times 3$-matrix ${\bf W}$
and ${\bf \Psi}^\dagger(\xi,\eta,w)\equiv
\overline{{\bf\Psi}^T(\xi,\eta,\overline{w})}$. With the use of
(\ref{UVlinsys}), (\ref{Wint}) the electrovacuum soliton solutions
have been constructed using the same dressing transformation ${\bf
\Psi}={\bf \chi}\cdot {\stackrel o {\bf \Psi}}$ with a more
particular, than (\ref{wsolvac}), soliton ansatz \cite{GA:1980}:
\begin{equation}\label{wsol}
{\bf \chi}={\bf I}+\sum\limits_{k=1}^{N}\displaystyle{{\bf
R}_k(\xi,\eta)\over w-w_k},\qquad {\bf \chi}^{-1}={\bf
I}+\sum\limits_{k=1}^{N}\displaystyle{{\bf S}_k(\xi,\eta)\over
w-\widetilde{w}_k} \end{equation}
where the poles of ${\bf \chi}$ are complex conjugated to the poles
of ${\bf \chi}^{-1}$, i.e. for each $k=1,2,\ldots, N$ we have
$\widetilde{w}_k=\overline{w_k}$. This leads to electrovacuum
generalization (which includes six real parameters per each
$w$-pole of ${\bf \chi}$) of the
Belinskii and Zakharov vacuum solitons with complex
conjugated poles, while a generalization of vacuum solitons with
real poles does not arise in this way. \footnote{Many electrovacuum
solutions which generalize vacuum solitons with real poles and on
some specially chosen backgrounds (e.g., on the Minkowski
background) can be constructed as the analytical continuations of
soliton solutions with complex poles in the space of their
constant parameters. This complex analytical continuation is quite
similar to the known one, which relates the "underextreme" and
"overextreme" parts of the Kerr - Newman family of solutions. Another
way for construction of such solutions is a direct integration of the
integral equation, which the spectral problem (\ref{UVlinsys}),
(\ref{Wint}) can be reduced to. This integral equation, derived
in \cite{GA:1985, GA:1988}, can be solved
and the solutions with arbitrarily large number of free constant
parameters can be constructed explicitly for rational
values of some special functional parameters in the
kernel of this equation, called as "monodromy data", and
expressed in terms of a number of arbitrary rational functions of
spectral parameter $w$ \cite{GA:1988, GA:1993b}.}

\section{Monodromy data of the solutions}
%
The analysis, suggested below, is based on the monodromy
transform approach, developed in \cite{GA:1985} - \cite{GA:1999}.
In this approach any local solutions of reduced vacuum or
electrovacuum Einstein equations is characterized unambiguously by
a finite set of functional parameters, which are functions of the
spectral parameter $w$ only and which admit a simple interpretation
as the monodromy data on the spectral plane for the fundamental
solution ${\bf \Psi}(\xi,\eta,w)$ of the spectral problem
(\ref{UVlinsys}), (\ref{Wint}), corresponding to a given solution of
the field equations under consideration.
In order to define the monodromy data, we need,
first of all, to fix some gauge freedom, existing in
(\ref{UVlinsys}), (\ref{Wint}).  For this we impose at some chosen
"initial" space-time point $(\xi_0,\eta_0)$\footnote{Actually, this
is a point in the orbit space of the space-time isometry group.} the
universal "normalization" conditions for the metric components
(say, ${\bf g}(\xi_0,\eta_0)=\epsilon_0\,
\mbox{diag\,}\{1,\epsilon\alpha_0^2\}$, where $\epsilon_0=\pm 1$),
which determine the value of ${\bf W}_o(w)\equiv {\bf
W}(\xi_0,\eta_0,w)$ as \begin{equation}\label{Wo} {\bf W}_o(w)= 4 i
(w-\beta_0)\left(\begin{array}{rrr}
0&1&0\\-1&0&0\\0&0&0\end{array}\right)+
 \left(\begin{array}{rrr}
-4\epsilon_0\epsilon\alpha_0^2&0&0\\0&-4\epsilon_0&0\\0&0&1
\end{array}\right)\end{equation}
where $\alpha_0=(\xi_0-\eta_0)/2 j$, $\beta_0=(\xi_0+\eta_0)/2$.
The normalization condition, used for the value of  ${\bf
\Psi}(\xi_0,\eta_0,w)$, which determines then the "normalized" value
of ${\bf K}(w)$, and the corresponding gauge transformations can be
chosen in the form
\begin{equation}\label{Psio}
\left.\begin{array}{l}
{\bf \Psi}(\xi_0,\eta_0,w)={\bf I}\\[3ex]
{\bf K}(w)={\bf W}_o(w)\\{}
\end{array}\hskip0ex\right\Vert\hskip0ex \begin{array}{l}
{\bf\Psi}\to {\bf \Psi}\cdot{\bf C}(w),\hskip3ex  {\bf C}(w)= {\bf
\Psi}^{-1}(\xi_0,\eta_0,w) \\[0.5ex]
\hskip-1ex\begin{array}{l}
 {\bf \Psi}\to {\bf A}\,{\bf \Psi}\,{\bf
A}^{-1},\\[0.5ex] {\bf W}\to ({\bf A}^\dagger)^{-1}\,{\bf W}\,
{\bf A}^{-1}, \end{array} \hskip0.5ex {\bf A}(w)=
\left(\hskip-1ex\begin{array}{ll}
SL(2,R)&\hskip-1.5ex\begin{array}{c} 0\\0\end{array} \\
\begin{array}{cc} a^3& a^4 \end{array}& \hskip-0.5ex 1
\end{array}\hskip-1ex\right)\end{array}
\end{equation}
where the $3\times 3$ - matrix ${\bf A}$ is constant. The
complex constants $a^3$ and $a^4$ change the additive constants in
the definitions of the components of a complex electromagnetic
potentials and $SL(2,R)$ part of ${\bf A}$ corresponds to a linear
transformation of the coordinates $x^3$, $x^4$ in (\ref{metrics}).

\begin{figure}[h] \epsfxsize=110mm
\centerline{\epsfbox{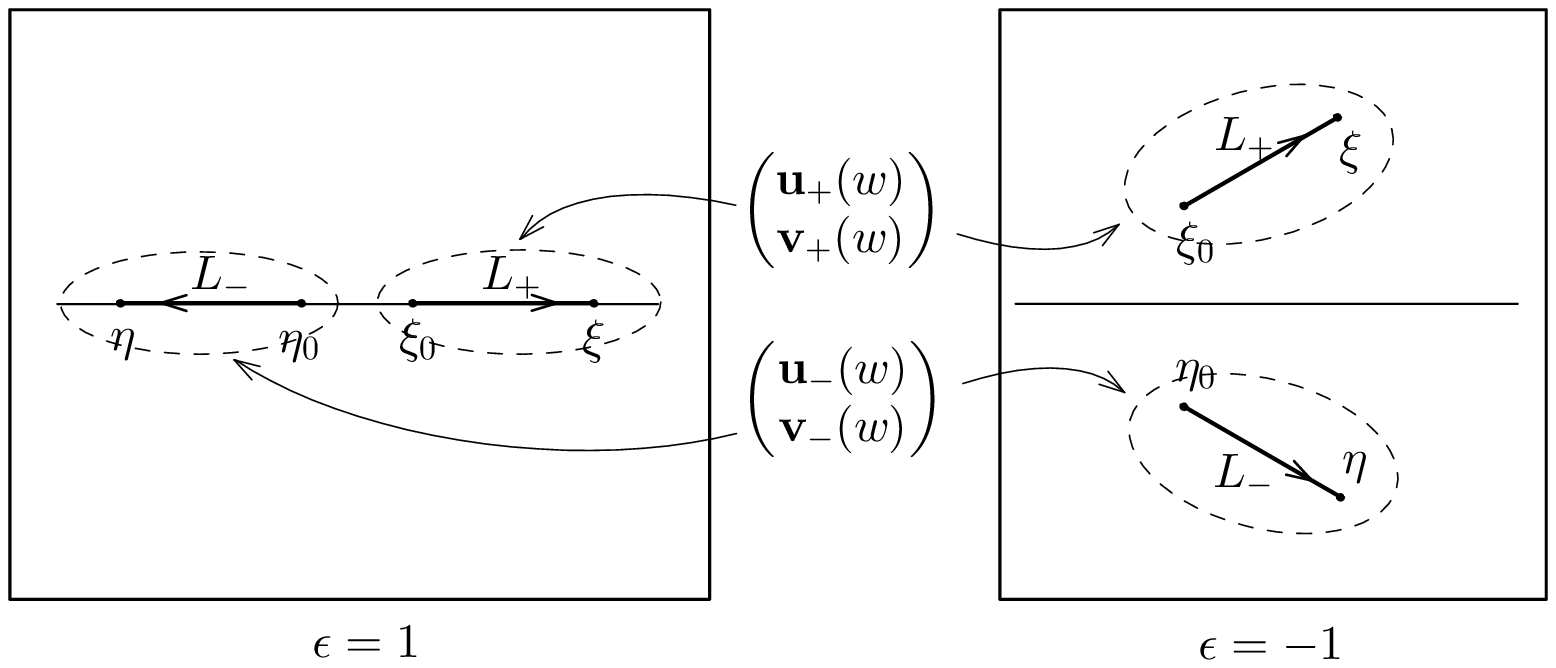}}
\caption{The singular points of ${\bf \Psi}$ and the structure of the
cut $L=L_++L_-$ on the spectral plane as well as the domains, where
the monodromy data functions ${\bf u}_\pm(w)$ and ${\bf v}_\pm(w)$
are defined and holomorphic, are shown here for the hyperbolic
($\epsilon=1$) and elliptic ($\epsilon=-1$) cases separately.}
\label{Fig_1} \end{figure}

A detail analysis \cite{GA:1985, GA:1988} of the analytical structure
on the spectral plane of the solutions of (\ref{UVlinsys}),
(\ref{Wint}) shows the existence of some universal properties of
${\bf \Psi}(\xi,\eta,w)$.\footnote{From now, the functions ${\bf
\Psi}$, ${\bf W}$, ${\bf K}$ are considered
in the gauges, fixed as in (\ref{Wo}), (\ref{Psio}).}
In particular, it is holomorphic
function of $w$ everywhere outside four algebraic branchpoints and
the cut $L=L_++L_-$ joining these points, as it is shown on Fig.\
\ref{Fig_1}.  It turns out, that the behaviour of ${\bf\Psi}$
near the branchpoints can be described by the monodromy
matrices ${\bf T}_\pm (w)$, which characterize the linear
transformations of ${\bf \Psi}$, continued analytically along the
paths $T_\pm$, rounding one of the branchpoints and joining different
edges of $L_+$ or $L_-$ respectively:
\begin{equation}\label{monodromy} {\bf \Psi}\quad{\stackrel
{T_\pm}\longrightarrow}\quad\widetilde{{\bf\Psi}}= {\bf \Psi}
\cdot{\bf T}_\pm(w),\qquad
{\bf T}_\pm(w)={\bf I}-2\displaystyle{{\bf l}_\pm(w)\otimes{\bf
k}_\pm(w)\over ({\bf l}_\pm(w)\cdot{\bf
k}_\pm(w))}.
\end{equation}
It is remarkable, that these matrices, satisfying the identities
${\bf T}_\pm^2(w)\equiv{\bf I}$, are independent of the space-time
coordinates $\xi$, $\eta$. The structure (\ref{monodromy}) allows
to express ${\bf T}_\pm$ in terms of the four complex projective
vectors ${\bf k}_\pm(w)$ and ${\bf l}_\pm(w)$, but it was found
in \cite{GA:1985}, that (\ref{Wint}) relate
unambiguously ${\bf l}_\pm(w)$ and ${\bf k}^\dagger_\pm(w)$ with the
same suffices. Therefore, both ${\bf T}_\pm$ are
determined completely by four scalar functions, which parametrize the
components of two projective vectors ${\bf k}_\pm(w)$:
\begin{equation}\label{mdata}
{\bf k}_\pm(w)=\{1\,,{\bf u}_\pm(w)\,,\,{\bf v}_\pm(w)\,\}
\end{equation}
The functions ${\bf u}_\pm (w)$, ${\bf v}_\pm (w)$, which domains
of holomorphicity are shown on \hbox{Fig.\ \ref{Fig_1}}, are called
as monodromy data. These data characterize unambiguously any local
solution of the field equations near some chosen "initial" or
"reference" point $(\xi_0,\eta_0)$, however, it is useful to keep in
mind, that for a given solution these data are dependent upon the
choice of this point. Note, that for pure vacuum case ${\bf
v}(w)\equiv 0$.

\section{Soliton generating transformations in terms of
the monodromy data}

Besides the definition of the monodromy data for any
local vacuum or electrovacuum solution, described in the previous
section, we recall here that these data enter explicitly
into the local structure of ${\bf \Psi}$ near the cuts $L_\pm$.
This structure can be described by the expressions \cite{GA:1988}:
\begin{equation}\label{local}
{\bf\Psi}=\lambda^{-1}_\pm\,{\bf
\psi}_\pm(\xi,\eta,w)\otimes {\bf k}_\pm(w)+{\bf
M}_\pm(\xi,\eta,w)\end{equation}
 where
$\lambda_+=\sqrt{(w-\xi)/(w-\xi_0)}$, $\lambda_-
=\sqrt{(w-\eta)/(w-\eta_0)}$ and the column-vectors ${\bf \psi}_\pm$,
the row-vectors ${\bf k}_\pm$ and the matrices ${\bf M}_\pm$ are
holomorphic at the points of the cut $L_+$ or $L_-$, respectively to
their suffices. In accordance with these expressions, the monodromy
data can be calculated from the branching parts of the components of
${\bf \Psi}$ on the cuts $L_\pm$.
We use now the expressions (\ref{local}) for calculation of the
transformations of the vectors ${\bf k}_\pm$ and therefore, of the
monodromy data, induced by the soliton generating
transformations.

It is easy to see, that any dressing transformation
${\bf\Psi}={\bf\chi}\cdot {\stackrel o {\bf \Psi}}(\xi,\eta,w)$ for
the normalized ${\bf \Psi}$ - functions reads
\begin{equation}\label{dress}
{\bf \Psi}={\bf A}\cdot{\bf\chi}
\cdot{\stackrel o {\bf \Psi}}\cdot{\bf\chi}_0^{-1}\cdot{\bf A}^{-1}
\end{equation}
where ${\stackrel o {\bf \Psi}}$ is the normalized background
solution, ${\bf \chi}_0(w)={\bf \chi}(\xi_0,\eta_0,w)$ and the
constant matrix ${\bf A}$ should be chosen afterwards for
normalization of the metric components and the matrix ${\bf W}$.
Let us consider now the expression (\ref{dress})
near the cuts $L_\pm$,
using there the local representations (\ref{local}) for ${\bf \Psi}$
and ${\stackrel o {\bf \Psi}}$. These expressions imply the following
transformations of the projective vectors ${\bf k}(w)$:
\vskip-0.75pc
\begin{equation}\label{kvectors}
{\bf k}_\pm(w)={\stackrel o {\bf k}}_\pm
(w)\cdot{\bf\chi}^{-1}_0(w)\cdot{\bf A}^{-1}. \end{equation}
\vskip-0.2pc\noindent
Now it is easy to calculate the monodromy data
${\bf u}_\pm(w)$, ${\bf v}_\pm(w)$ for solitons, as the
ratios of the components of the projective vectors (\ref{kvectors}),
using one of the soliton ansatz (\ref{wsolvac}) or (\ref{wsol}).
The result can be presented in the linear-fractional form
(for vacuum background ${\stackrel o {\bf v}}_\pm(w)\equiv 0$
and for vacuum solitons ${\bf v}_\pm(w)\equiv 0$):
$${\bf u}_\pm (w)= \displaystyle{{\cal U}_0 +{\cal U}_1\,
{\stackrel o {\bf u}}_\pm(w)+{\cal U}_2\,  {\stackrel o {\bf
v}}_\pm(w) \over {\cal Q}_0 +{\cal Q}_1\, {\stackrel o {\bf
u}}_\pm(w)+{\cal Q}_2\, {\stackrel o {\bf v}}_\pm(w)},\hskip1ex {\bf
v}_\pm(w)=\displaystyle{{\cal V}_0 +{\cal V}_1\, {\stackrel o {\bf
u}}_\pm(w)+{\cal V}_2\, {\stackrel o {\bf v}}_\pm(w)\over {\cal Q}_0
+{\cal Q}_1\, {\stackrel o {\bf u}}_\pm(w)+{\cal Q}_2\, {\stackrel o
{\bf v}}_\pm(w)},$$ where all coefficients are polynomials in $w$,
which orders do not exceed the number of w-solitons $N$ (or $2 N$ in
the Belinskii and Zakharov formalism).

\section*{Acknowledgments}
This work was partly supported by the British Engineering and
Physical Sciences Research Council and the Russian Foundation for
Basic Research Grants 99-01-01150, 99-02-18415.

\end{document}